\documentclass{PoS}

\newcommand{\ov}{\overline v_0}
\newcommand{\ove}{\overline v_{05}}

\title{Gauge-Higgs Unification on the Lattice}

\ShortTitle{Gauge-Higgs Unification on the Lattice}

\author{\speaker{Nikos Irges}%
 \\
Department of Physics, National Technical University of Athens\\
Zografou Campus, GR-15780 Athens Greece\\
         E-mail: \email{irges@mail.ntua.gr}}

\author{Francesco Knechtli and Kyoko Yoneyama\\
        Department of Physics, Bergische Universit\"at Wuppertal\\
        Gaussstr. 20, D-42119 Wuppertal, Germany\\
        E-mail: \email{knechtli@physik.uni-wuppertal.de,yoneyama@physik.uni-wuppertal.de}}

\abstract{The simplest Gauge-Higgs Unification model is a five-dimensional $SU(2)$ gauge theory
compactified on the $S^1/Z_2$ orbifold, such that on the four-dimensional boundaries of space-time there 
is an unbroken $U(1)$ symmetry and a complex scalar, the latter identified with the Higgs boson.
Perturbatively the $U(1)$ remains spontaneously unbroken. Earlier lattice Monte Carlo simulations revealed however that 
the spontaneous breaking of the $U(1)$ does occur at the non-perturbative level. Here, we 
verify the Monte Carlo result via an analytical lattice Mean-Field expansion. \begin{flushright} WUB/12-19 \end{flushright}}

\FullConference{The 30th International Symposium on Lattice Field Theory\\
		 June 24 - 29,  2012\\
		 Cairns, Australia}

\begin{document}

\section{Introduction}

The discovery of a Higgs-like boson at the LHC raises several questions, one of them being an understanding
of the origin of the associated mechanism of Spontaneous Symmetry Breaking (SSB). 
One of the most exciting explanations is a possible connection of the mechanism to the existence of
an extra dimension \cite{Hoso}. 
In this, "Gauge-Higgs Unification" (GHU) scenario, the Higgs boson comes from some of the extra dimensional components
of the five-dimensional gauge field. 
The model we will discuss here is an $SU(2)$ gauge theory in five dimensions with the fifth dimension
compactified on the $S^1/Z_2$ orbifold. The embedding of the orbifold action in the gauge field $A_M^A$ 
where $M=\mu, 5$ is a Lorentz index and $A=1,2,3$ is a gauge index, is such that
$A_\mu^3$ with $\mu=0,1,2,3$ and $A_5^{1,2}$ are even, and all other components are odd. 
The latter is the "Higgs", a complex scalar in the fundamental representation of the 
unbroken $U(1)$ boundary symmetry. 
This is the simplest, prototype model.
Its generalization to an $SU(3)$ bulk symmetry leaves an $SU(2)\times U(1)$ symmetry
on the boundary with two complex doublet Higgs fields, in the fundamental representation
of the $SU(2)$ factor. We believe that if the mechanism of SSB is present in the simpler system,
it will be generically present also in the more complicated cases, so we study it first
in the $SU(2)$ model. We also take the point of view that SSB is driven by
pure gauge dynamics, a fact that has been observed in earlier Monte Carlo simulations \cite{MC1}.
In fact, a perturbative 1-loop computation of the Coleman-Weinberg potential does not yield SSB
in the pure gauge system at infinite cut-off \cite{Kubo}. With a finite cut-off it is possible in principle to have SSB \cite{NFM},
however it is not possible to prove this in the perturbative context since the quantum theory is non-renormalizable.
All this points to the necessity for developing a non-perturbative analytical tool which can probe the system 
near its bulk phase transition \cite{Creutz}, where a scaling regime with suppressed cut-off effects might exist.
Such a formalism has been developed in \cite{MF1,MF2} and it is an expansion in fluctuations
around a Mean-Field (MF) background \cite{DZ}. 
There is serious evidence that the MF expansion describes the non-perturbative system 
with periodic boundary conditions faithfully \cite{FrancAM}, so we will now apply it to the orbifold model as well.

The parameters of the model are the dimensionless five-dimensional lattice coupling $\beta$,
the anisotropy $\gamma$ and the lattice size which is set by $(T,L)$ points along the $(\mu=0, \mu=1,2,3)$ 
directions and $N_5+1$ points along the extra dimension.
$L$ will be taken always large enough so that physics does not depend on it. Typically this 
happens when $L\ge 200$ approximately. As the lattice action, the Wilson plaquette action is used,
with the orbifold boundary conditions and the anisotropy appropriately implemented in it \cite{MF2}.

\section{The Mean-Field expansion}

The first step in the MF formalism is to trade the gauge links $U$ of the lattice with unconstrained complex
variables $v$ and a set of Lagrange multipliers $h$ that ensure that the memory of the gauge nature of the links is not lost.
Then, the gauge links can be integrated out.
The resulting effective action is then minimized with respect to the left over degrees of freedom $v$ and $h$.
The saddle point solution defines the MF background. 
We will be considering lattices wich are isotropic in the $\mu$ directions and have an anisotropy
$\gamma$ along the fifth dimension. 
Consequently, the background is $\ov(n_5)$ along four-dimensional hyperplanes and $\ove(n_5+1/2)$ along
the fifth dimension. Here $n_5$ is the discrete label of extra dimensional points on the lattice.
The phases of the system are defined as follows:
\begin{itemize}
\item Confined phase: $\ov(n_5), \ove(n_5+1/2) = 0$.
\item Layered phase: $\ov(n_5) \ne 0$, $\ove(n_5+1/2) = 0$.
\item Deconfined phase: $\ov(n_5), \ove(n_5+1/2) \ne 0$.
\end{itemize}
The above solutions are found by solving numerically the non-linear algebraic equations that reflect the saddle point of the path integral.
The boundary conditions ensure that this anisotropic solution is not an artifact, since translational invariance
is broken along the fifth dimension by the presence of the boundaries.

We parametrize the fluctuations around the MF background as
\begin{eqnarray}
v&=&v_0+iv_A\sigma^A, v_{0,A}\in C \nonumber\\
h&=&h_0+ih_A\sigma^A, h_{0,A}\in C \, .\nonumber
\end{eqnarray}
These are introduced in the formalism by the substitution 
\begin{eqnarray}
&& v \longrightarrow \ov + v \nonumber\\
&& h \longrightarrow \bar h_0 + h \nonumber
\end{eqnarray}
in the path integral and performing a derivative expansion on the effective action and on the gauge invariant observable
$\cal O$. 

\begin{figure}\centering
  \resizebox{6cm}{!}{\includegraphics[angle=0]{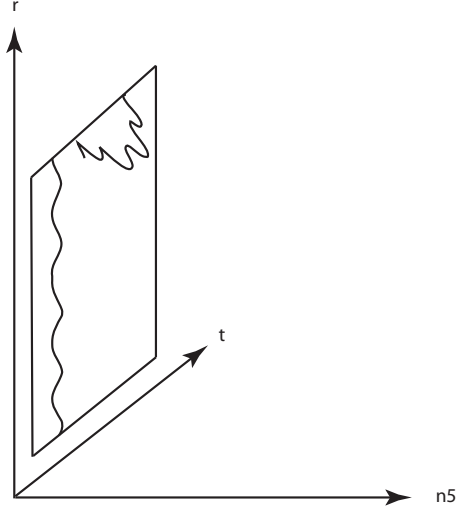}}
  \caption{Contributions to the static potential on the boundary of the orbifold:
  gauge boson exchange and self energy.}
  \label{Wilson}
\end{figure}

The second derivative part of this expansion defines the lattice propagator 
\begin{equation}
{K}_{M'M''}^{-1} = {K}_{M'M''}^{-1}(p',n'_5,M',\alpha' ;p'',n''_5,M'',\alpha'')\, ,
\end{equation}
with $p=(p_0,p_k)$ the four-dimensional lattice momenta and $\alpha=0,A$ a gauge index.
The quantity
\begin{equation}
\langle {\cal O} \rangle  =
{\cal O}[{\ov}] + \frac{1}{2}{\rm tr} 
\left\{\frac{\delta^2{\cal O}}{\delta v^2}\Biggr|_{\ov} K^{-1}\right\} \label{correction}
\end{equation}
defines an observable to first order in the fluctuations. The time-dependent correlator $C(t)$ is then defined as
\begin{equation}
C (t) = \langle {\cal O} (t_0+t){\cal O} (t_0)\rangle -
\langle {\cal O} (t_0+t)\rangle  \langle {\cal O} (t_0)\rangle\, ,
\end{equation}
with $t_0$ an arbitrary initial time and the mass of the associated ground state is extracted from
\begin{equation}
m = \lim_{t\to \infty} \ln \frac{C (t)}{C (t-1)}\, .\label{mass}
\end{equation}
The main observable of our interest will be the Wilson Loop ${\cal O}_W(r,t)$ of length $r$ along one of the dimensions on the boundary.
The static potential extracted as
\begin{equation}
t\to \infty : \hskip .5cm {\rm e}^{-V(r) t} \simeq \; <{\cal O}_W>  
\end{equation}
contains the key information about SSB: if the static potential is of a four-dimensional Yukawa form 
with the Yukawa mass identified with the mass of the boundary gauge boson, then we
have a spontaneously broken $U(1)$ on the boundary.
In this case, we are entitled to call this gauge boson a "$Z$ boson".
To leading order in the MF expansion, the two types of gauge boson exchange diagrams 
appearing on fig. \ref{Wilson} dominate the Wilson Loop.

%
%

Specializing to our $SU(2)$ model, for the static potential, we arrive at the result \cite{MF2}:
\begin{eqnarray}
&&V(r)=-\log({\ov(0)^2})-\frac{1}{2}\frac{1}{L^3}
\frac{1}{(\ov(0))^2}\sum_{p_k'}\nonumber\\
&& \Biggl\{\frac{1}{3}\sum_k\Bigl[ 2\cos{(p_k'r)}+2\Bigr]
{K}^{-1} \left((0,p_k'),0,0,0;(0,p_k'),0,0,0\right)\nonumber\\
&+& \frac{1}{3}\sum_k\Bigl[ 2\cos{(p_k'r)}-2\Bigr]
{K}^{-1}\left((0,p_k'),0,0,3;(0,p_k'),0,0,3\right)\Biggr\}\, .
\label{potb}
\end{eqnarray}
There is a similar expression for the Wilson Loop parallel to the above, sitting in the middle of the fifth dimension.

The other observable that we will use is the one with scalar quantum numbers, corresponding to the Higgs field.
It is basically a pair of Polyakov Loops (projected on the orbifold) along the fifth dimension separated in the time direction, exchanging a gauge boson.
The computation of this diagram yields the Higgs correlator $C_H(t)$
(for details see \cite{MF2}):
\begin{equation}
C_H(t) = \frac{8}{L^3 T}(P_0^{(0)})^2  \Pi^{(1)}_{\langle 1,1\rangle}(0,0) \,,
\end{equation}
where $P_0^{(0)}$ is the Polyakov loop evaluated on the background and
$\Pi^{(1)}_{\langle 1,1\rangle}(0,0)$ is
\newpage
\begin{eqnarray}
\Pi^{(1)}_{\langle 1,1\rangle}(0,0) &=& 
2\sum_{p_0'}\cos{p_0't} \sum_{n_5',n_5''} \sum_{r=0}^{N_5-1}\frac{\delta_{n_5', r}}{\ov(r+1/2)}\cdot \nonumber\\
&& {K}^{(-1)}((p_0',{\vec 0}),n_5',5,0;(p_0',{\vec 0}),n_5'',5,0)
\sum_{r=0}^{N_5-1}\frac{\delta_{n_5'', r}}{\ov(r+1/2)}\, .\label{Higgs}
\end{eqnarray}
Then the Higgs mass is extracted according to eq. (\ref{mass}).

\section{Spontaneous Symmetry Breaking}

In principle one could attempt to massage further the result eq. (\ref{potb}) and in particular consider its small lattice spacing expansion.
We will not attempt such a task here, instead we will compute it numerically and fit it to the various possible
forms that the static potential could assume. Furthermore, we will present cases where only a four-dimensional Yukawa fit is possible.
Having the Standard Model Higgs mechanism in mind, we will compute the quantity
\begin{equation}
\rho_{HZ} = \frac{m_H}{m_Z}\label{rho}
\end{equation}
with $m_H$ the mass of the complex scalar and $m_Z$ the mass of the boundary $U(1)$ gauge boson,
extracted from the Yukawa fit.
This quantity is infinite in the absence of SSB and finite when there is SSB.
In our framework it depends on three parameters: $\beta$, $\gamma$ and $N_5$.
We fix $\gamma=0.55$ and fix $\beta$ so that $F_1=m_H R$ (with $R$ the length of the interval - fifth dimension)
is constant and follow its $N_5$-dependence.
In fact, it is possible to extract from the static potential not only the Yukawa mass $m_Z$ but also
the mass of the first excited state. Such a state, from the point of view of models beyond the Standard
Model is typically called a $Z'$. Thus, we can define a similar to eq. (\ref{rho}) quantity
\begin{equation}
\rho_{HZ'} = \frac{m_H}{m_{Z'}}\label{rhoprime}.
\end{equation}
Clearly, by computing both eqs. (\ref{rho}) and (\ref{rhoprime}) and fixing to a definite value one of them,
would give us a prediction for the other.
The recent result from the LHC motivates us to fix (approximately)
\begin{equation}
\rho_{HZ} = 1.3875\label{SMrho}
\end{equation}
which then leaves us with a definite prediction for $m_{Z'}$. The caveat here of course is that such a
process would have to be performed along a Line of Constant Physics (LCP) \cite{LCP}.
Here we will only present data to show that $\rho_{HZ}$ is finite and therefore argue that there is
a non-perturbative dynamical mechanism of SSB in the pure gauge system. We use the term dynamical in order to stress that
the gauge boson becomes massive without introducing by hand a vacuum expectation value beyond that
of the MF background. 

\begin{figure}\centering
  \resizebox{10cm}{!}{\includegraphics[angle=0]{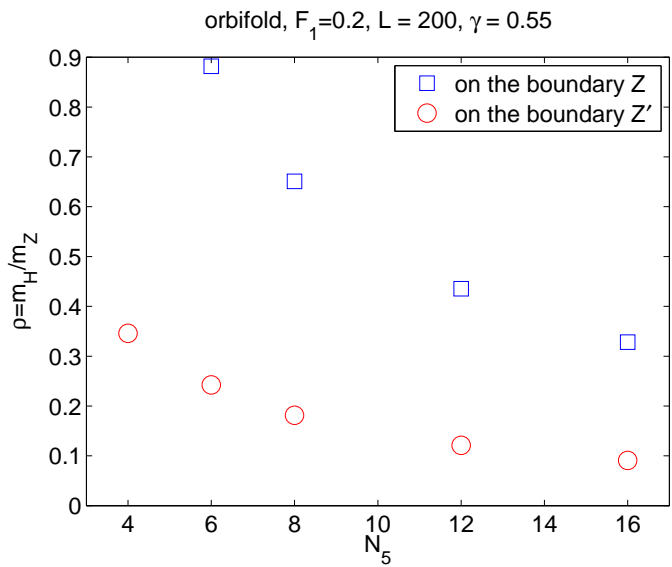}}
  \caption{The ratio of the Higgs to the $Z$ and $Z'$ boson masses
  in the mean-field extracted from the static potential on the boundary.}
  \label{f_rho_g0p55}
\end{figure}

On fig. \ref{f_rho_g0p55} we show our main result regarding SSB in the lattice $SU(2)$ orbifold model in the
MF expansion, for $\gamma=0.55$ and $F_1=0.2$. Evidently the order parameter that signals SSB is not infinite
for a wide range of $N_5$ values, consistent with earlier lattice Monte Carlo results \cite{MC1}.
Moreover, preliminary results show that it is possible to get for
$\rho_{HZ}$ the value of eq. (\ref{SMrho}), an analysis that will be presented in \cite{LCP}.

%
%

\section{Conclusions}

We computed the Higgs to $Z$-boson mass ratio in a five-dimensional $SU(2)$ gauge theory
regularized on an anisotropic lattice, with the anisotropy pointing along the fifth-dimension.
The method is that of an analytical Mean-Field expansion around a non-trivial background,
which is evidently a good approximation to the non-perturbative theory in five (or higher) dimensions.
We computed this quantity in the vicinity of the bulk phase transition.
Contrary to the analogous calculation in the perturbative regime, we find that there is
spontaneous symmetry breaking of the boundary $U(1)$ symmetry already
in the pure gauge theory.  The breaking is dynamical, since no Higgs vacuum expectation value
is introduced and is consistent with results from Monte Carlo investigations.

{\bf Acknowledgments.} K. Y. is supported by the Marie Curie Initial
  Training Network STRONGnet.
  STRONGnet is funded by the European Union under Grant Agreement number
  238353 (ITN STRONGnet).
  N. I. thanks the Alexander von Humboldt Foundation for support.
  N. I. was partially supported by the NTUA research program PEBE 2010.

\end{document}